\PassOptionsToPackage{numbers}{natbib}
\documentclass{winnower}
\usepackage{graphicx}
\usepackage{subcaption}

\begin{document}

\title{Using Intel Optane Devices for In-situ Data Staging in HPC Workflows}

\author{Pradeep Subedi, Philip E. Davis, J. J. Villalobos, Ivan Rodero, and Manish Parashar\\
\tt{\{pradeep.subedi, philip.e.davis, jj.villalobos, irodero, parashar\} @rutgers.edu}}

\affil{Rutgers Discovery Informatics Institute $(RDI^2)$, Rutgers University}


\date{}

\maketitle

\begin{abstract}
Emerging nonvolatile memory technologies (NVRAM) offer alternatives to hard drives that are persistent, while providing similar latencies to DRAM. Intel recently released the Optane drive, which features 3D XPoint memory technology. This device can be deployed as a SSD or as persistent memory.
In this paper, we provide a performance comparison between Optane (SSD DC4800X) and NVMe (SSD DC3700) drives as block devices. We study the performance from two perspectives: 1) Benchmarking of drives using FIO workloads, and 2) Assessing the impact of using Optane over NVMe within the DataSpaces framework for in-memory data staging to support in-situ scientific workflows.

\end{abstract}

\section{Introduction}
Large-scale scientific simulation workflows are generating increasing amounts of data at very high rates, and this data has to be transported and analyzed before scientific insights can be obtained. These simulation workflows are also becoming increasingly complex, and are composed of coupled applications and data processing components that need to interact and exchange data. As a result, data management, transportation, processing and analysis costs are quickly dominating and limiting the potential impacts of these advanced simulations \cite{prabhakar2011provisioning}. 

In-situ/In-transit data processing approaches, which utilize in-memory data staging, are garnering popularity due to their ability to use compute and storage resources to process data close to where they are generated \cite{zhang2012enabling}. Data staging reduces the end-to-end runtime of high-performance scientific workflows by enabling the efficient use of compute node DRAM for I/O management \cite{docan2012dataspaces}. However, data volume is growing to unprecedented scale and DRAM offers limited capacity as well as minimal persistence for data. As a result, modern supercomputers are increasingly incorporating additional levels of memory hierarchy, ranging from on-node SSDs to shared burst buffer appliances \cite{liu2012role}.

The adoption of non-volatile memory devices such as solid-state drives (SSDs) in various HPC deployments, such as the Summit supercomputer at Oak Ridge National Laboratory, has presented new opportunities to design how data is offloaded from simulation applications and consumed by analysis applications. SSDs offer several benefits over traditional hard drives due to their lower data-access latency, lower power consumption, and increased stability. Furthermore, these lower costs and larger capacities as compared to DRAM make SSD an attractive candidate to compose an intermediate data storage level to address the performance and latency gaps between DRAM and high latency storage systems. However, additional complexities associated with managing placement of data across multiple layers of the memory hierarchy and its access by multiple concurrently executing tasks present significant challenges.

Intel and Micron recently released Optane SSD based on 3D-XPoint memory technology. It is claimed to be faster and more endurable than NAND flash based memory devices \cite{foong2016storage}, \cite{coughlin2016crossing}, \cite{micheloni20163d}. In addition to being usable as a block device, it has the capability to be used as a DRAM extension, i.e., the Optane drive will be seen as a part of the main memory. Although using the byte-addressability feature of the Optane device might simplify the storage hierarchy by transparently inserting the Optane into the addressable memory space, this also limits the ability to use Optane as a shared device, such as burst buffer, which is seen by multiple nodes as persistent storage.

To this end, we explore the performance characteristics of Optane SSD as a block device and compare it with NVMe SSD storage. We also integrate the Optane SSD into the memory hierarchy of the DataSpaces framework\cite{docan2012dataspaces} and study its impact on the overall response time of applications reading and writing data to data-staging servers.

\section{Comparing performance of Optane vs NVMe SSD}
In our evaluation tests for characterizing the read/write throughput of Optane Drives, we used an Intel Xeon node, which has 28 Intel(R) Xeon(R) CPU E5-2660 v4 @ 2.00GHz cores. This node is equipped with 128 GB of DRAM. The node runs CentOS 7.4.1708 x86\_64 and has a standard set of development libraries along with FIO-benchmark installed. The Optane Drive and NVMe SSD used for testing were DC4800X and DC3700, respectively. 

\begin{figure}
    \centering
    \begin{subfigure}[b]{0.48\textwidth}
    	\vspace{10pt}
        \includegraphics[width=\textwidth]{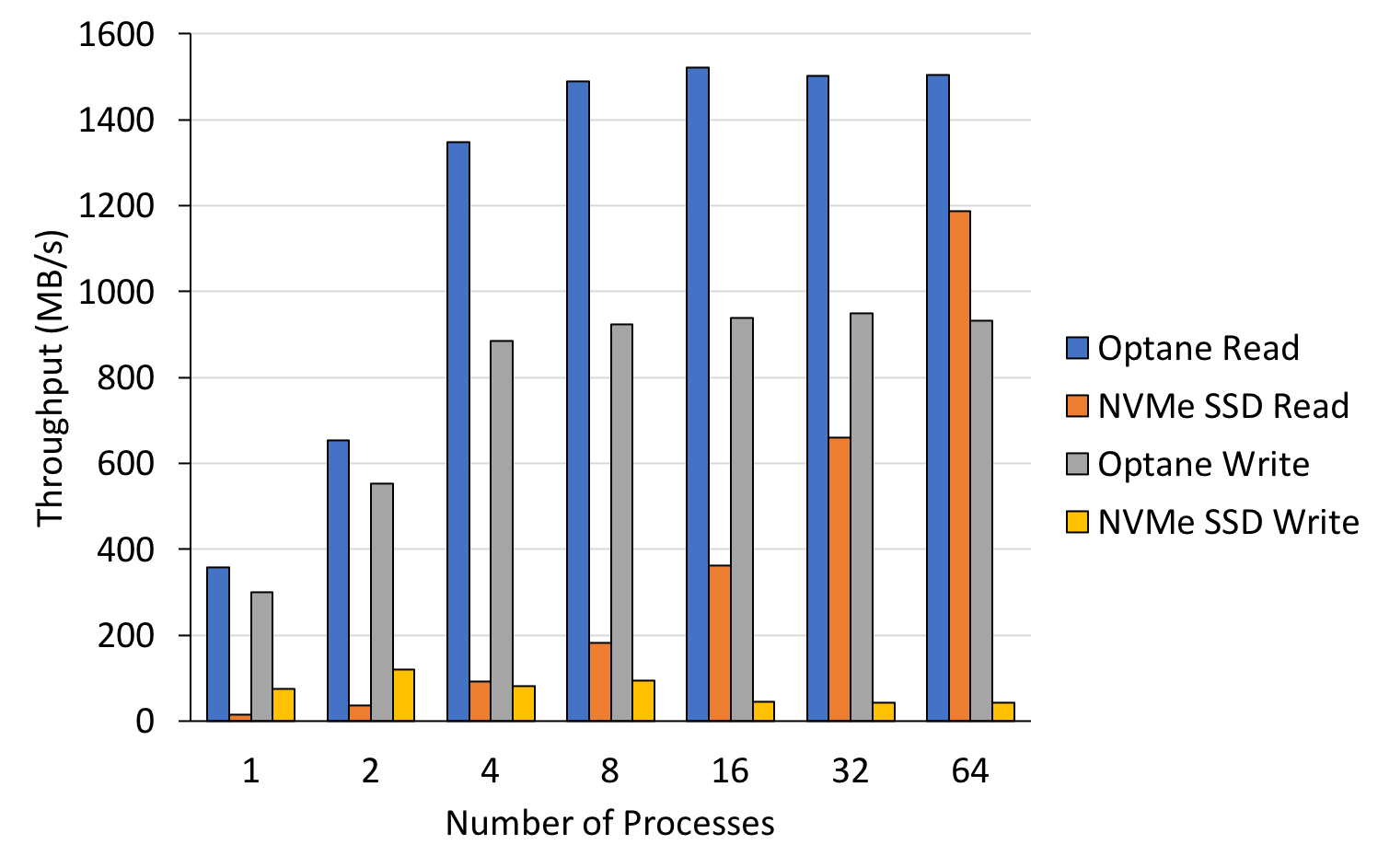}
        \caption{Random Access}
        \label{fig:rand_rw_4k}
    \end{subfigure}
    \begin{subfigure}[b]{0.48\textwidth}
    	\vspace{10pt}
        \includegraphics[width=\textwidth]{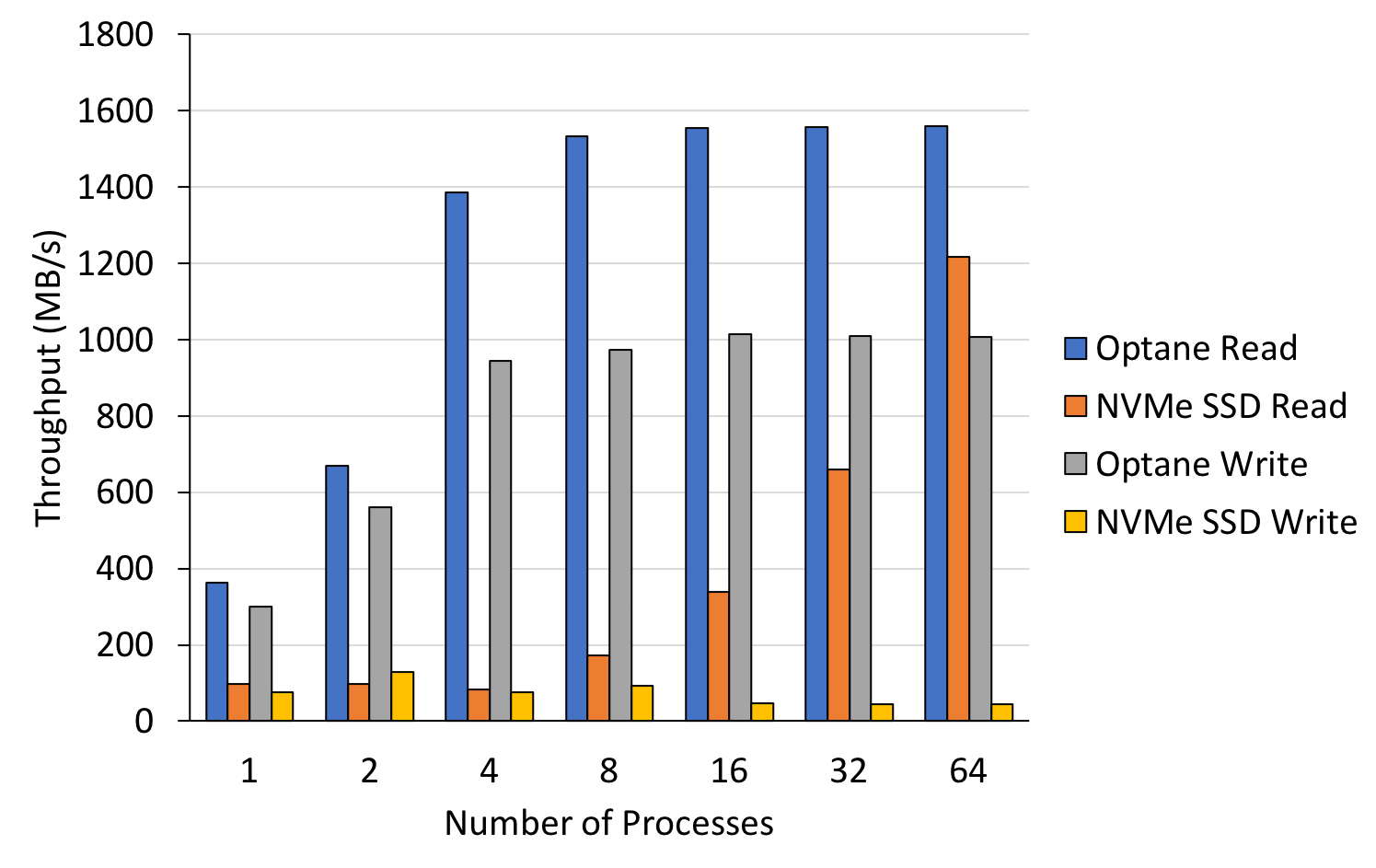}
        \caption{Sequential Access}
        \label{fig:seq_rw_4k}
    \end{subfigure}
    \caption{Throughput for Optane and NVMe SSD for reads and writes with 4KB transfer size.}\label{fig:rand_read_write}
\end{figure}

Figure \ref{fig:rand_read_write} shows the comparison of throughput for Intel Optane Drive and Intel NVMe SSD for a varying number of processes. FIO-benchmark was used to issue random and sequential read/write requests to the block devices. The sample $fio$ script below shows the test with direct write to device enabled, with queue depth of 1 and transfer size of 4KB for random read workload.

\vspace{10pt}
\text{\tt{fio --filename=\$ssd\_dir --direct=1 --rw=randrw --refill\_buffers --norandommap}}

\text{\tt{--randrepeat=0 --ioengine=libaio --bs=4k --rwmixread=100 --iodepth=1 --numjobs=1}}

\text{\tt{--runtime=300  --name=4ktest --size=128G --output=4krandomRead\_QD1.op}}\vspace{10pt}

We used similar scripts for all of our tests with varying number of processes, transfer size and random/sequential access patterns.
In the case of 4KB transfer size, it was observed that increasing the number of processes reading data from the Optane/NVMe SSD increases the aggregate throughput seen by the application. From Figure \ref{fig:rand_read_write}, we can clearly see that Optane outperforms NVMe SSD in terms of both read and write performance. An interesting observation that we made in the course of our evaluation was that write performance of Optane is very similar to its read performance. Optane reached a sequential write throughput of up to 1005MB/sec, while NVMe SSD only reached 128MB/sec for a queue depth of 1. 
In the case of NVMe, the max read throughput was observed to be only up to 1214 MB/sec, in contrast to Optane's 1559MB/sec for 64 process with a queue depth of 1. It is worth noting that by increasing transfer size to 64KB and changing queue depth to 128, we were able to achieve a random read throughput of 2530MB/sec and random write throughput of 2171MB/sec with Optane, while NVMe SSD achieved read and write throughputs of 1217MB/sec and 864MB/sec. It is clearly evident from these results that Optane can significantly outperform NVMe SSDs and can directly impact the perceived latencies for I/O intensive applications.

\begin{figure}
    \centering
    \begin{subfigure}[b]{0.49\textwidth}
    	\vspace{10pt}
        \includegraphics[width=\textwidth]{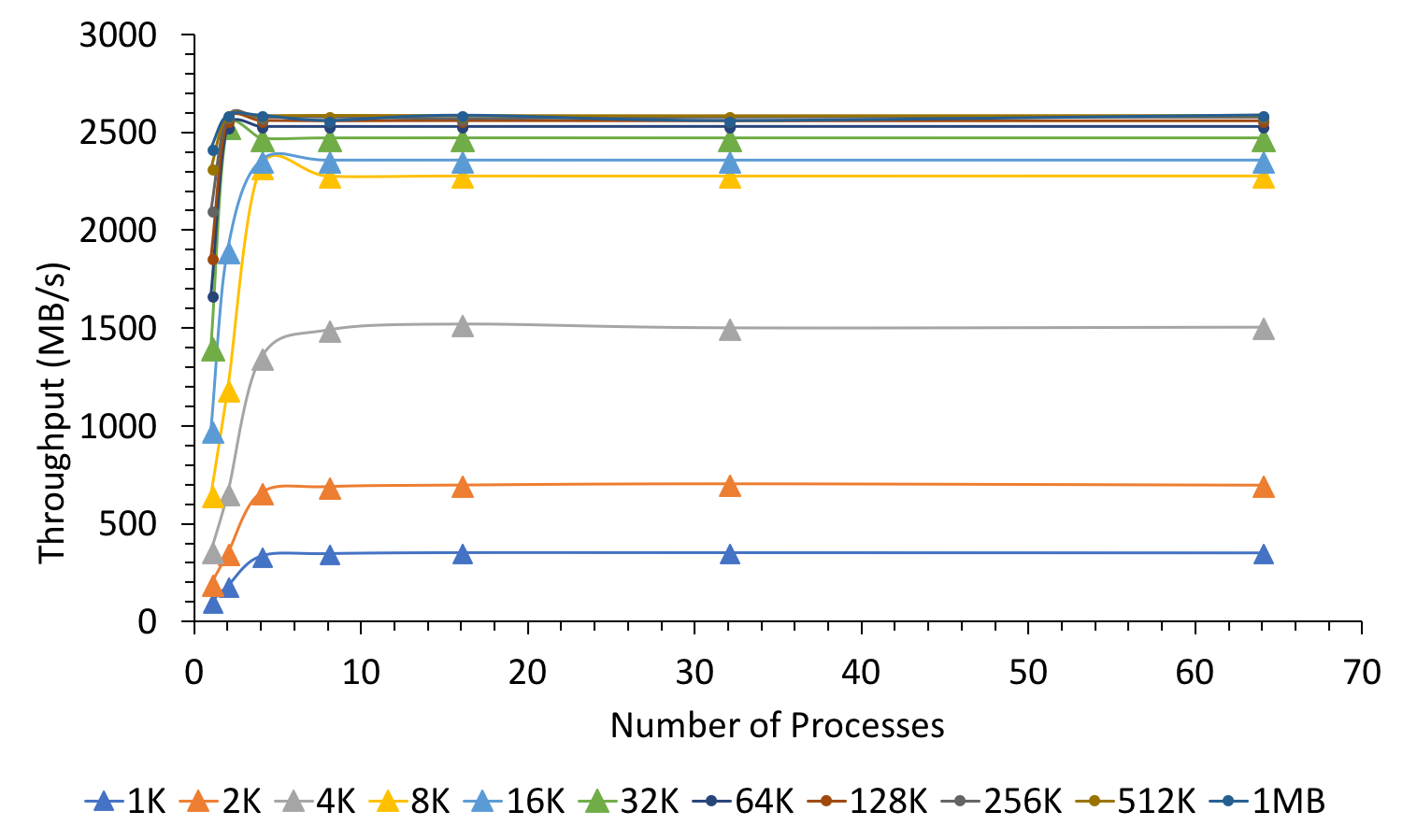}
        \caption{Optane }
        \label{fig:optane_rand_read}
    \end{subfigure}
    \begin{subfigure}[b]{0.49\textwidth}
    	\vspace{10pt}
        \includegraphics[width=\textwidth]{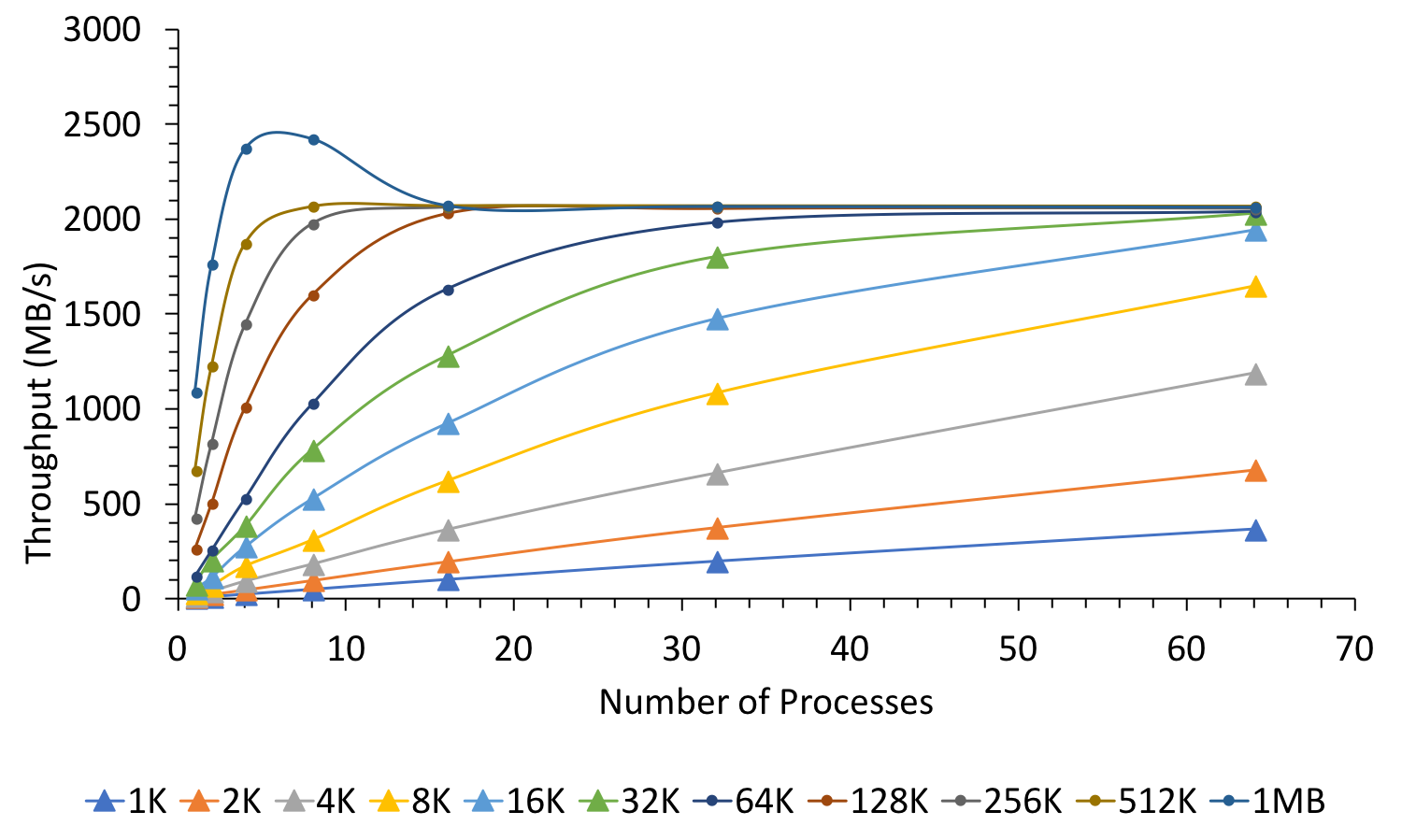}
        \caption{NVMe SSD}
        \label{fig:ssd_rand_read}
    \end{subfigure}
    \caption{Random Read Throughput for Optane and NVMe SSD for transfer size vs. Number of Processes }\label{fig:rand_read}
\end{figure}

\begin{figure}
    \centering
    \begin{subfigure}[b]{0.49\textwidth}
    	\vspace{10pt}
        \includegraphics[width=\textwidth]{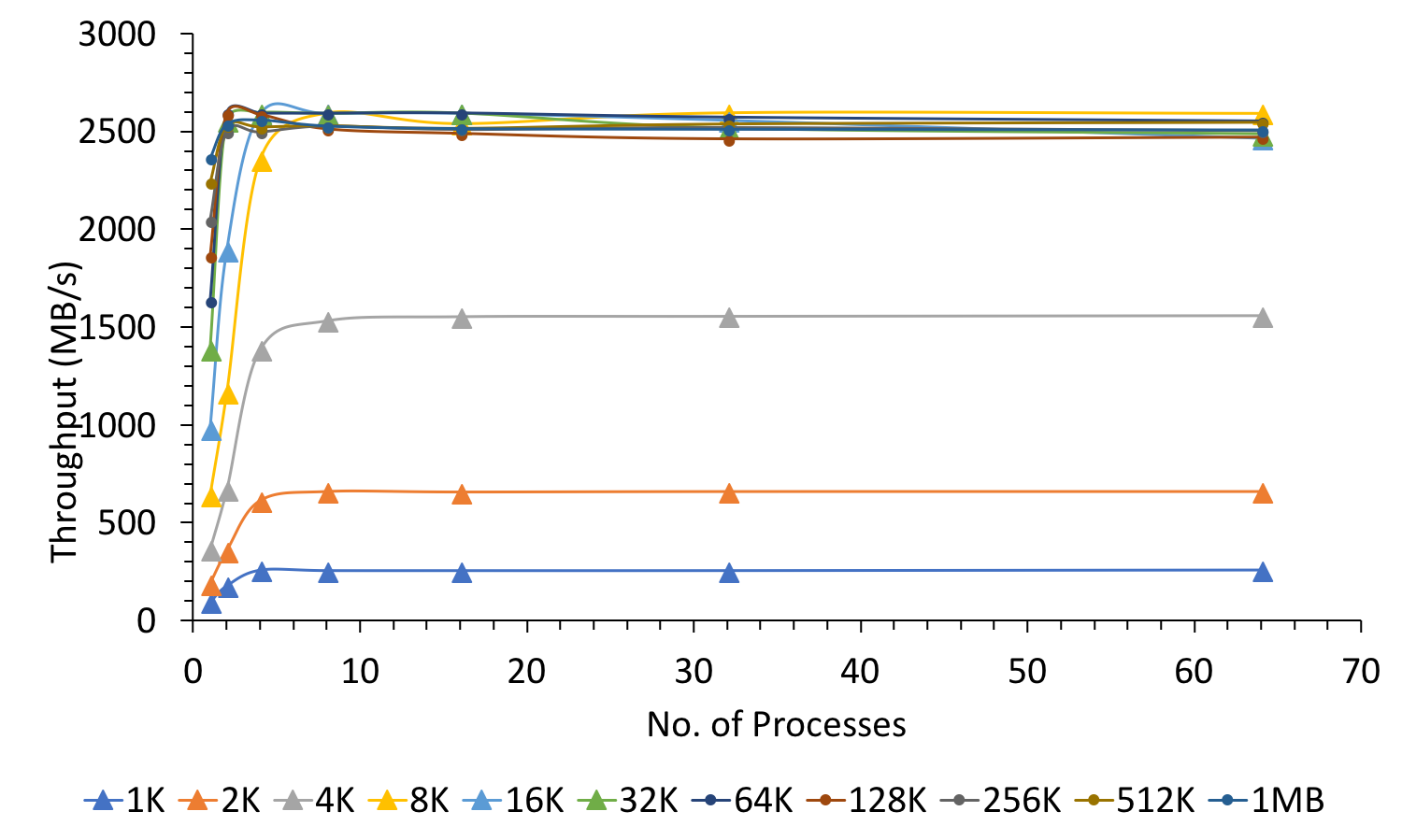}
        \caption{Optane}
        \label{fig:optane_seq_read}
    \end{subfigure}
    \begin{subfigure}[b]{0.49\textwidth}
    	\vspace{10pt}
        \includegraphics[width=\textwidth]{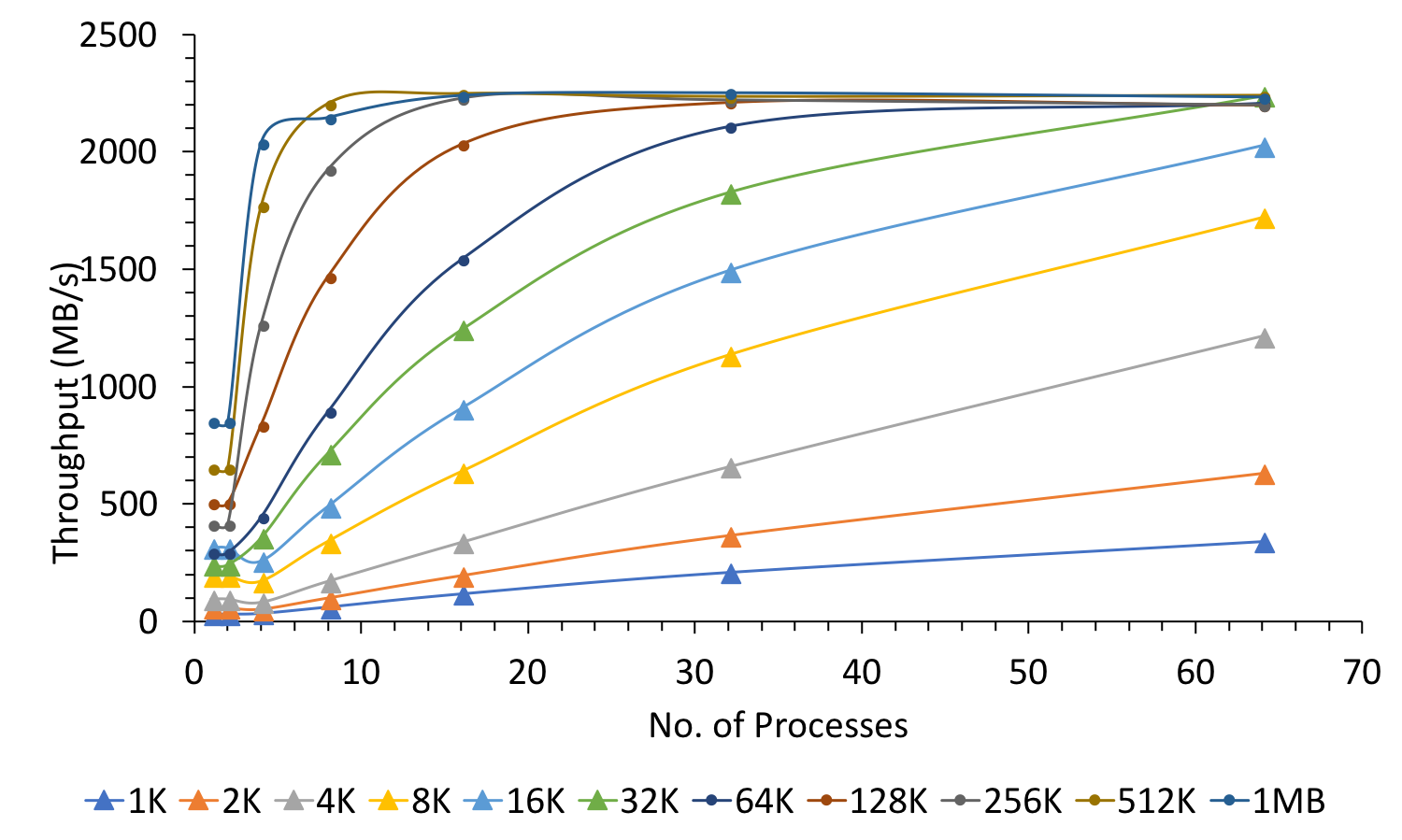}
        \caption{NVMe SSD}
        \label{fig:ssd_seq_read}
    \end{subfigure}
    \caption{Sequential Read Throughput for Optane and NVMe SSD for transfer size vs. Number of Processes }\label{fig:seq_read}
\end{figure}

To better understand the random and sequential read access patterns' impact on both drive types, we varied the data transfer size from 1 KB to 16 MB, with the number of processes ranging from 1 to 64 in Figure \ref{fig:rand_read} and \ref{fig:seq_read}. It was observed that increasing the transfer size for each request can drastically affect the application's perceived throughput. With 256 KB transfer size, Optane was able to achieve its near maximum throughput, while even 64 processes could not utilize the full bandwidth offered by Optane drive when transfer size was set to 4 KB. A similar trend was seen for NVMe SSD as well. This leads us to believe that Optane drives have higher read throughput capacity and for the applications to fully utilize the available bandwidth they must either either multiple threads requesting data at the same time or larger I/O queue depth. Otherwise, the drives will be underutilized. In comparion with NVMe, Optane demonstrated better read throughput for all test cases.

\begin{figure}
    \centering
    \begin{subfigure}[b]{0.49\textwidth}
    	\vspace{10pt}
        \includegraphics[width=\textwidth]{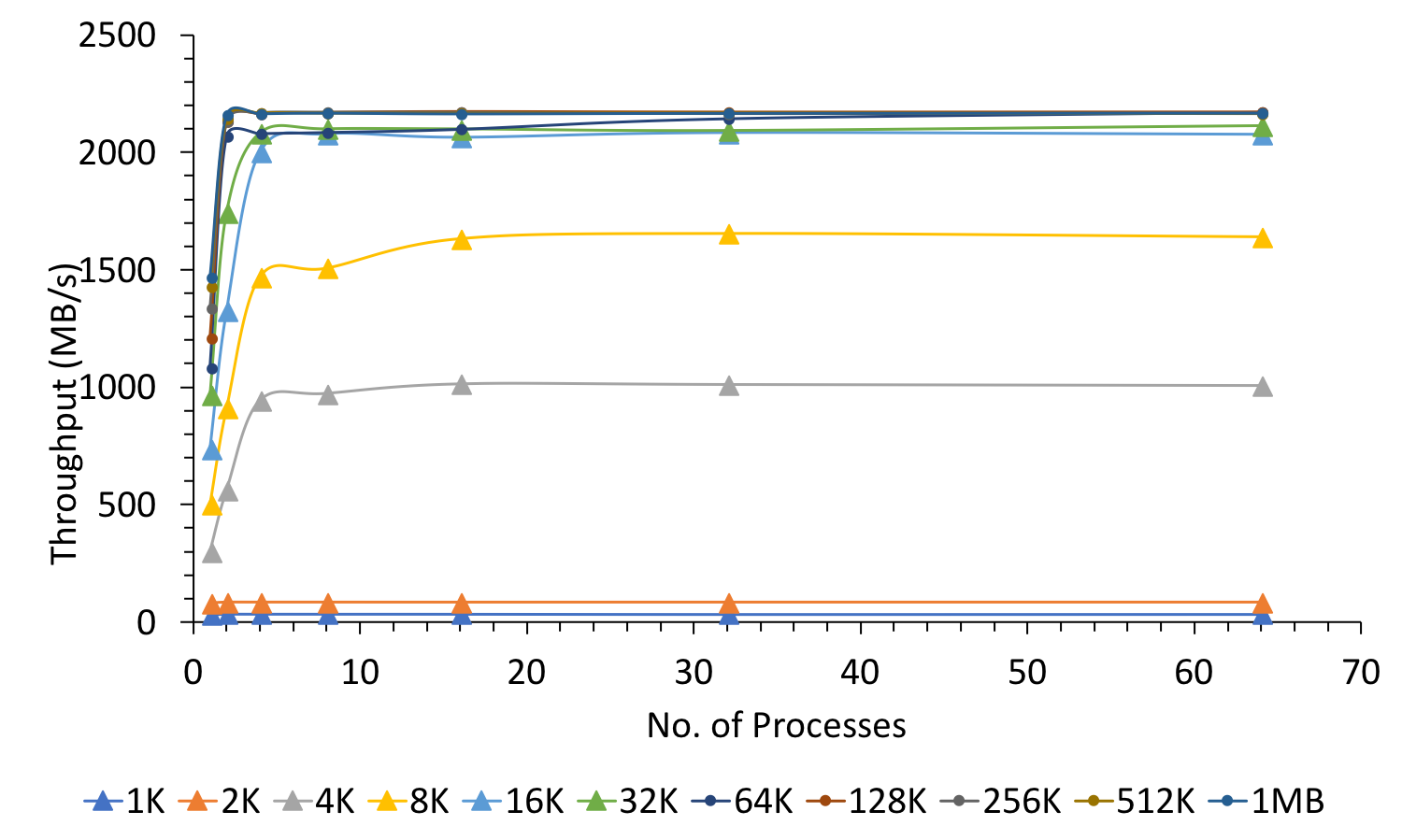}
        \caption{Optane }
        \label{fig:optane_seq_write}
    \end{subfigure}
    \begin{subfigure}[b]{0.48\textwidth}
    	\vspace{10pt}
        \includegraphics[width=\textwidth]{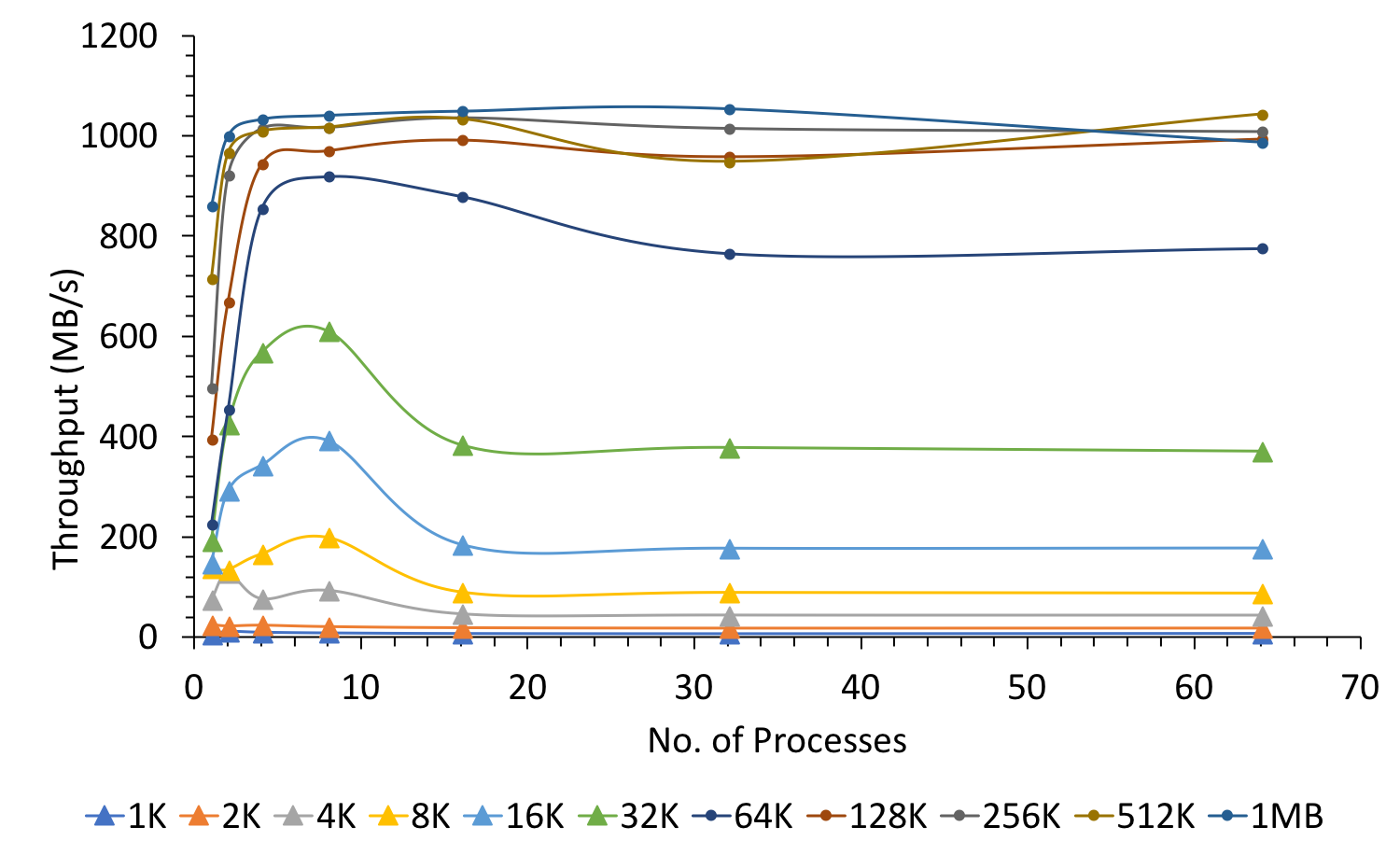}
        \caption{NVMe SSD}
        \label{fig:ssd_seq_write}
    \end{subfigure}
    \caption{Sequential Write Throughput for Optane and NVMe SSD for transfer size vs. Number of Processes }\label{fig:seq_write}
\end{figure}

\begin{figure}
    \centering
    \begin{subfigure}[b]{0.49\textwidth}
    	\vspace{10pt}
        \includegraphics[width=\textwidth]{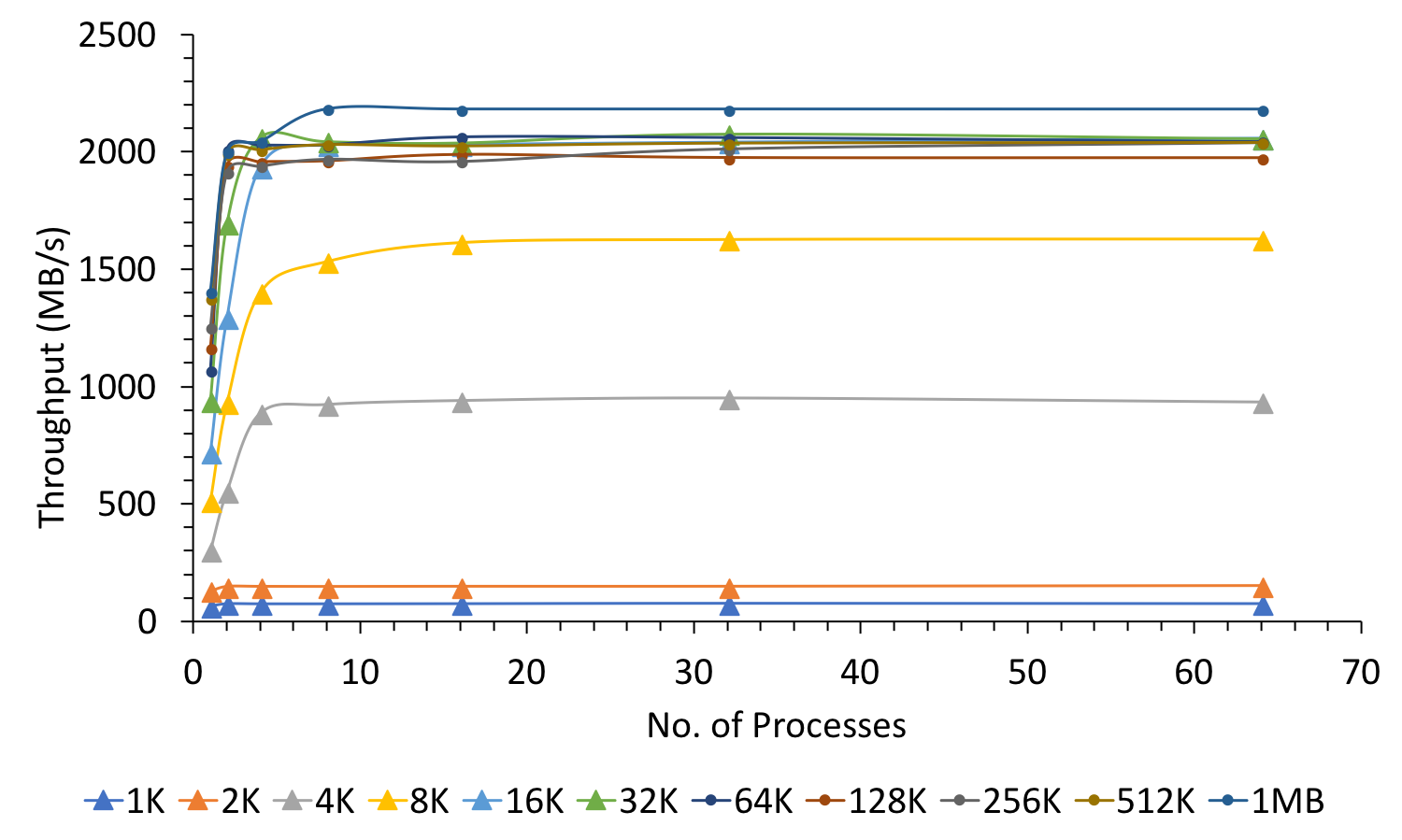}
        \caption{Optane }
        \label{fig:optane_rand_write}
    \end{subfigure}
    \begin{subfigure}[b]{0.48\textwidth}
    	\vspace{10pt}
        \includegraphics[width=\textwidth]{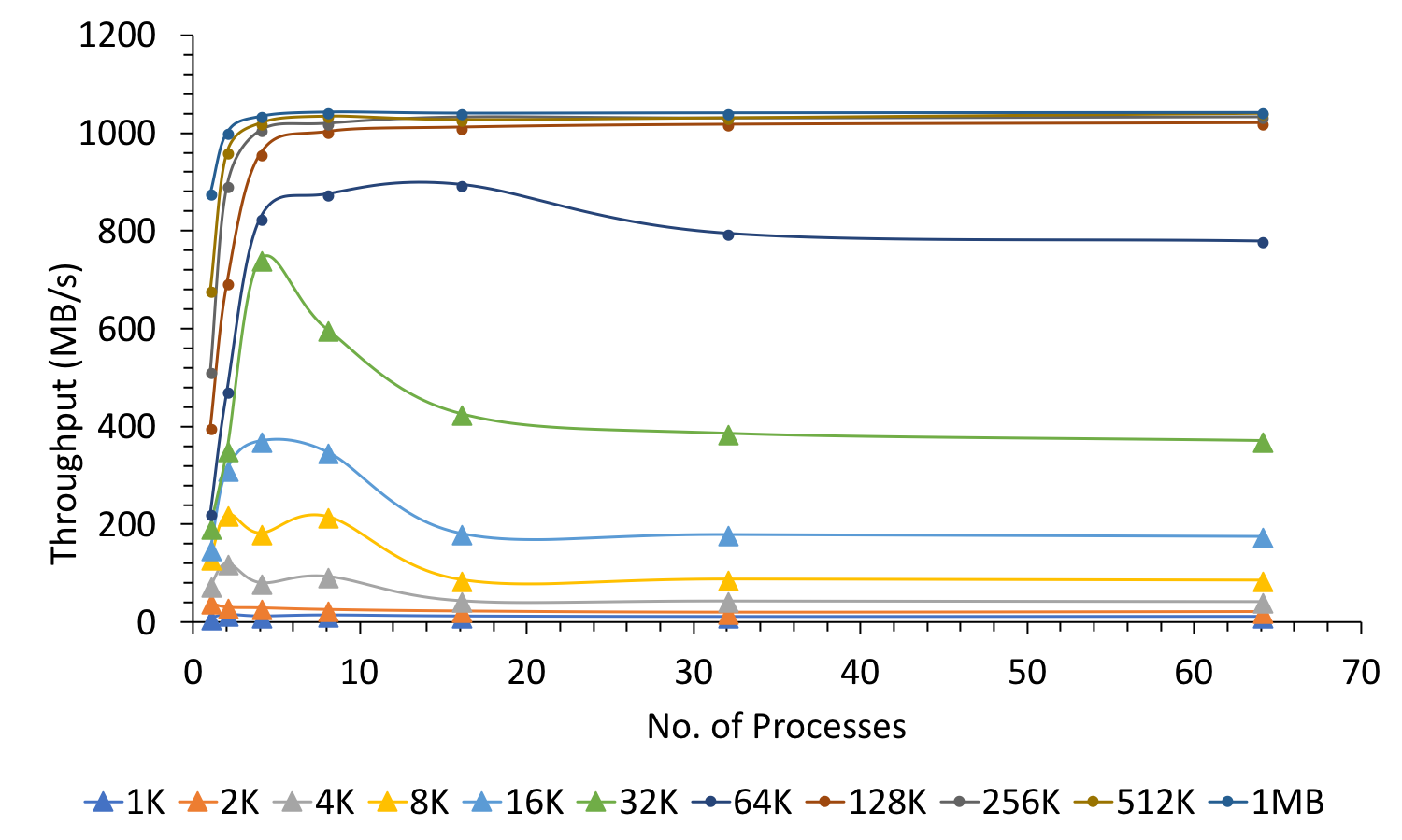}
        \caption{NVMe SSD}
        \label{fig:ssd_rand_write}
    \end{subfigure}
    \caption{Random Write Throughput for Optane and NVMe SSD for transfer size vs. Number of Processes }\label{fig:rand_write}
\end{figure}

A similar set of tests were also conducted for write-only workloads. From Figures \ref{fig:seq_write} and \ref{fig:rand_write}, we can see that Optane drives have almost twice the write throughput of NVMe SSD for both random and sequential writes. While write throughput remains fairly constant for Optane drives, in the case of NVMe, the write throughput scales linearly up to 8 processes, but then overall throughput starts to decline. We think this decline is due to the write interference from multiple processes, since all writes are going to the same drive. From these results, we see that issuing requests in the size of 128KB transfer size for both sequential and random writes can help us to better utilize these drives by achieving the maximum write throughput.

\section{Enabling Staging across the Deep Memory Hierarchy using DataSpaces}

DataSpaces is a programming system targeted at current large-scale systems and designed to support dynamic interaction and coordination patterns between scientific applications. DataSpaces essentially provides a semantically specialized shared-space abstraction using a set of dedicated staging processes. This abstraction derives from the tuple-space model and can be associatively accessed by the interacting applications of a simulation workflow. DataSpaces also provides services including distributed in-memory associative object store, scalable messaging, as well as runtime mapping and scheduling of online data analysis operations. DataSpaces is currently being used by production coupled scientific simulation workflow on large-scale supercomputers.

\begin{figure}
    \centering
    \begin{subfigure}[b]{0.48\textwidth}
    	\vspace{10pt}
        \includegraphics[width=\textwidth]{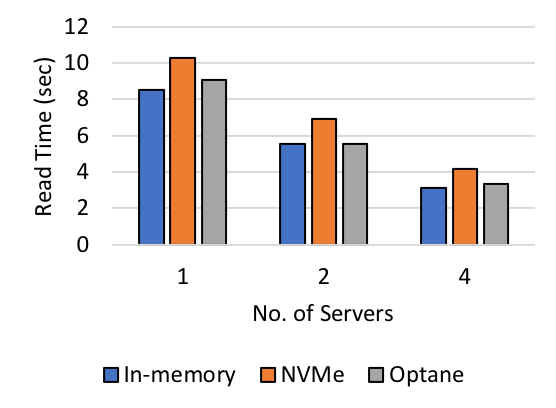}
        \caption{Read Response Time for 64 readers reading data in parallel}
        \label{fig:optane_rand_write}
    \end{subfigure}
    \begin{subfigure}[b]{0.469\textwidth}
    	\vspace{10pt}
        \includegraphics[width=\textwidth]{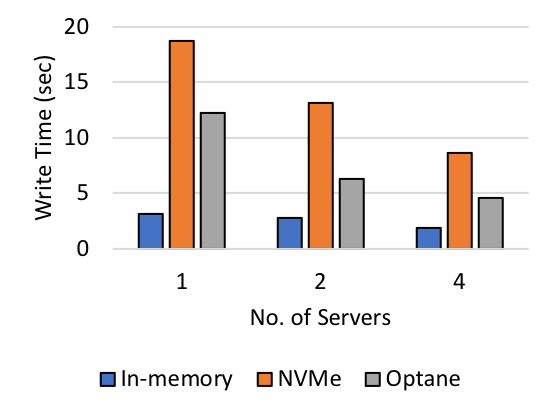}
        \caption{Write Response Time for 64 writers writing data in parallel}
        \label{fig:ssd_rand_write}
    \end{subfigure}
    \caption{Response Times in Seconds for Reads/Writes to varying DataSpaces servers. All of the servers write/read to/from the same drive. Total data transferred in each direction is 4GB (Strong Scaling).}\label{fig:strong_serv}
\end{figure}

As part of evaluating the impact of using block devices for data staging, we added extensions in the DataSpaces framework to support writing the staged data to Optane/NVMe SSDs. In particular, we modified the way data is stored in DataSpaces servers. When a writer application wants to make data available to readers, it sends that data to the DataSpaces server for storage. To enable deep memory hierarchy, we modified the DataSpaces code such that during initialization a memory mapped file is created. This memory mapped file is backed by block device (Optane or NVMe). Any data that is being written to DataSpaces servers is re-directed to this memory mapped file. Once the data transfer is completed, the data is flushed to drive and other servers are notified of data being available for read requests. Since the file is memory mapped, all reads are treated as reading from memory, and the operating system internally handles page swapping to and from the drive for the requested data.

For evaluation purposes, we used 3 Intel Xeon Phi x200 (KNL - 72 cores) node, which has Intel(R) Xeon Phi(TM) CPU 7290 @ 1.50GHz cores and one Intel Omni-Path NIC card. Each/cat node is equipped with 192 GB of DRAM. The nodes are individually dedicated to the role of readers, writers or servers. For example in the 64 readers, 64 writers and 4 servers case, one node runs all readers, another node runs all writers and the final node runs 4 DataSpaces servers. In all of the test cases, the reader and writer applications were run for 10 time steps i.e, 64 writers wrote a total of 4GB data and 64 readers read a total of 4 GB data for 10 iterations. The results reported in this section are the average of these 10 time-step/iteration runs.

\begin{figure}
    \centering
    \begin{subfigure}[b]{0.48\textwidth}
    	\vspace{10pt}
        \includegraphics[width=\textwidth]{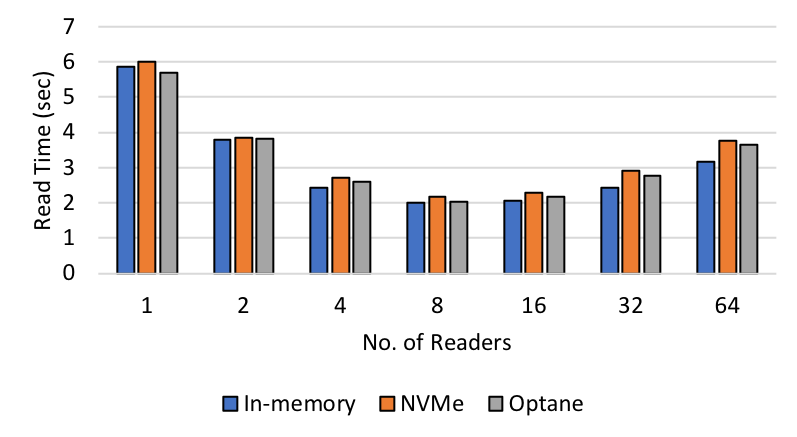}
        \caption{Read Response Time for readers reading data in parallel. Total number of writers is 64.}
        \label{fig:reader_strong}
    \end{subfigure}
    \begin{subfigure}[b]{0.486\textwidth}
    	\vspace{10pt}
        \includegraphics[width=\textwidth]{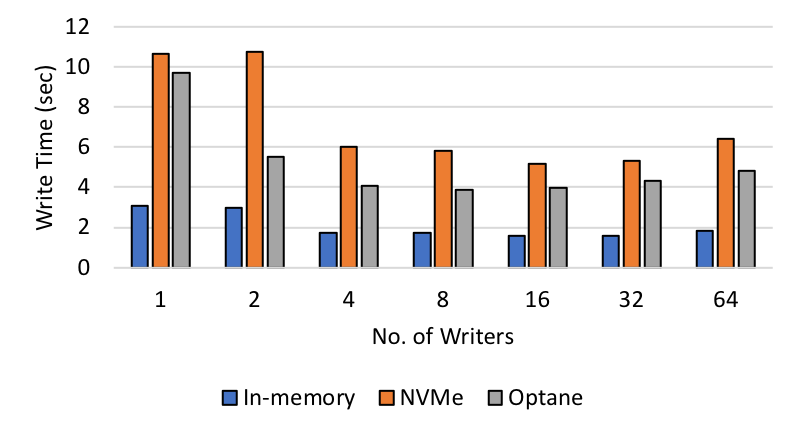}
        \caption{Write Response Time for writers writing data in parallel. Total number of readers is 64.}
        \label{fig:writer_strong}
    \end{subfigure}
    \caption{Response Times in Seconds for Reads/Writes to varying readers and writers. Total number of DataSpaces servers in 4 and Total data transferred in each direction is 4GB (Strong Scaling). }\label{fig:strong_read_write}
\end{figure}

Figure \ref{fig:strong_serv} shows the read and write response time for 64 readers and 64 writers with the number of data staging servers varying. The total data transferred in each direction in each time step is 4GB in this case. We can clearly see that for writers, Optane Drives provide better response time than NVMe SSDs, while both of them are slower than DRAM. The reason behind this is the overhead of copying data to and from DRAM to the drive. The important thing to note is that running multiple servers on the same node rather than single server to process same workflow reduces the write/read response time. This benefit comes from the parallelism achieved with multiple servers and the Optane drive's ability to serve the write/read request without significant delay. Note that, since the file is memory mapped, the operating system internally handles the loading of data to memory, which leads to a fluctuation of read response time for Optane vs. NVMe. 

We also varied the number of readers/writers and also performed the tests for both weak and strong scaling. In strong scaling, the total data is kept constant, whereas in weak scaling the data per client (reader/writer) is kept constant, thereby scaling the aggregate data size. Figure \ref{fig:strong_read_write} shows the read and write response time for strong scaling experiments. In Figure \ref{fig:reader_strong}, the number of writers is kept to 64, while readers range from 1 to 64. Since total data transferred is constant, the amount of data being read per reader increases when fewer readers are used. As the readers increase, the parallelism increases, but increasing readers beyond 8 or 16 increases the synchronization costs among readers and results in smaller requests being issued to the DataSpaces servers. This leads to an increase in read response time. Figure \ref{fig:writer_strong} shows a variation of the number of writers, while readers were held to 64. We observed that 16 writers seemed to maintain a balance between parallelism and write request size being issued to DataSpaces servers. For a single writer case, we did not observe a significant improvement over the NVMe use case and we believe this is a limitation of a single KNL core as it is not able to drive Optane to its full capacity. With an increase in the number of writer cores, we can easily see that performance improvement over NVMe.

\begin{figure}
    \centering
    \begin{subfigure}[b]{0.48\textwidth}
    	\vspace{10pt}
        \includegraphics[width=\textwidth]{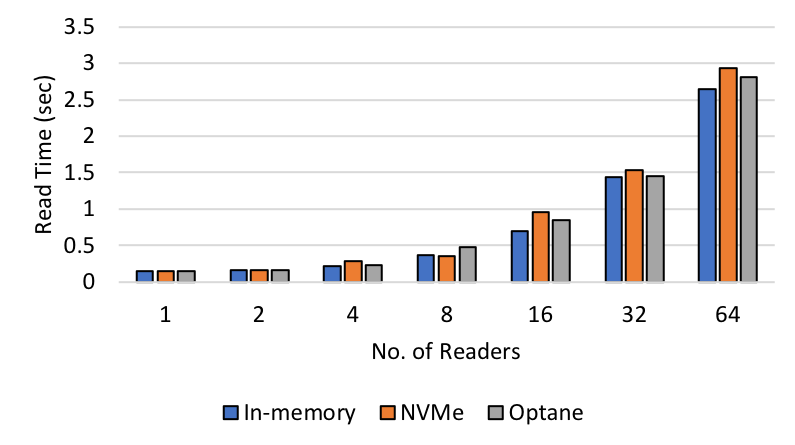}
        \caption{Read Response Time for readers reading data in parallel. Total number of writers is 64.}
        \label{fig:reader_weak}
    \end{subfigure}
    \begin{subfigure}[b]{0.486\textwidth}
    	\vspace{10pt}
        \includegraphics[width=\textwidth]{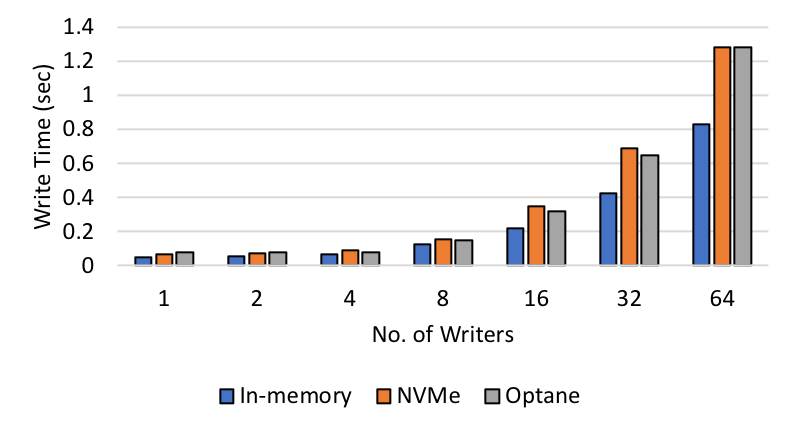}
        \caption{Write Response Time for writers writing data in parallel. Total number of readers is 64.}
        \label{fig:writer_weak}
    \end{subfigure}
    \caption{Response Times in Seconds for Reads/Writes to varying readers and writers. Total number of DataSpaces servers in 4 and Data Transferred per read/writer in each direction is 8 MB (Weak Scaling). }\label{fig:weak_read_write}
\end{figure}

It should be noted that, although Optane provides significant read/write throughput improvement over NVMe as seen on FIO tests, the amount of performance improvement perceived by the application such as DataSpaces depends upon how the files are written to the drives. In the case of DataSpaces, files were memory mapped which hid most of the disk read access latency. Thus, the performance observed for DRAM, NVMe and Optane were somewhat similar for read times in Figures \ref{fig:strong_read_write} and \ref{fig:weak_read_write}.

Figure \ref{fig:weak_read_write} shows the read and write response time for weak scaling experiments. Here, data per reader/writer is kept at 8 MB. Since the amount of data being transferred increases with the increase in readers/writers, the read/write response time increases. The Optane vs NVMe performance results follow the same pattern as seen in earlier experiments of strong scaling.

\section{Discussion}
In-situ workflows can generate a huge amount of data. Ideally, all the data exchanges between components could be kept in DRAM, but that is not feasible with large scientific applications. Distributed file systems are generally not practical for in-situ coupling either, due to their relatively poor performance when compared to the rate at which HPC nodes generate and consume data. Data staging is thus well positioned to take advantage of node-local resources that can be provided to applications as a service, especially as a resource that expands the storage capacity of the staging node without reducing the data access performance too much. SSD has historically filled this role, but the access latencies of SSD are disadvantageous for some common access patterns. Optane is a quantitative leap forward in this regard, allowing more performant in-situ workflows. The capability of Optane as a byte-addressable device (as a part of memory) significantly reduces the coding efforts to integrate the block devices into data staging frameworks. While Optane reduces the coding efforts, the data movement between DRAM and Optane in the byte addressable configuration is hidden to users. This might adversely affect the research efforts to optimize data transfers between DRAM and block/byte-addressable devices.

Although some internals are hidden from researchers view, the performance of OS-integrated byte-addressing can act as a baseline for researchers who are integrating block devices into the memory. Optane drives present a growing opportunity for in-situ workflows because they are seen as part of memory and provide low latency along with persistent storage. An increase in the persistent memory capacity has the potential to trigger significant shifts in HPC application architecture from in-transit data analysis techniques to in-situ data analysis. 

Additionally, this report shows that, for some data access patterns, I/O bandwidth exceeds the ability of a single core to move data. This needs to be considered as a part of the scalable data staging framework design. The ideal data staging framework for in-situ and in-transit workflows would be multi-threaded, where multiple threads are issuing read/write requests to Optane, so that I/O bandwith is fully utilized to reduce the data transfer latency from DRAM to Optane as a block device.

\section{Conclusion}
In this report, we have studied the performance of Optane drives and compared it with that of NVMe SSDs. We also evaluated the impact of deep memory hierarchy in the data-staging framework. Given the benefits of Optane over NVMe, Optane is likely to replace NVMe SSDs in HPC nodes. While Optane drives can expand the DRAM memory capacity when used in a byte-addressable configuration, a single application with smaller queue depth might not be enough to fully drive the Optane's capacity leading to lower utilization of available bandwidth. To fully utilize the high bandwidth capacity of Optane, these drives need to be shared by multi-threaded or multi-process applications. In the data staging use case, the use of $mmap$ hides the benefit of Optane over NVMe for read workloads, while Optane showed clear advantage over NVMe SSDs for write workloads. In conclusion, we can say that while Optane drives provide promising results for write intensive applications in HPC workflows, implementation specifics such as memory mapping might hide those advantages for various use cases such as burst-buffers and data staging and design decisions should be made to utilize the high I/O bandwidth provided by Optane drives.

\bibliographystyle{abbrvnat}
\bibliography{winnower_template}

\end{document}